\documentclass[useAMS,usenatbib]{mn2e}

\usepackage{url,times,graphicx,amsmath,amsfonts,amssymb,aas_macros,color,epsfig,epstopdf}

\newcommand{\mhalo}{M$_{\rm halo}$}
\newcommand{\mstar}{M$_{\rm *}$}
\newcommand{\mHI}{M$_{\rm HI}$}
\newcommand{\mgas}{M$_{\rm gas}$}

\newcommand{\mb}{M$_{\rm b}$}
\newcommand{\lcdm}{$\Lambda$CDM}
\newcommand{\msun}{M$_\odot$}

\newcommand{\vmax}{V$_{\rm max}$}
\newcommand{\vrot}{V$_{\rm rot}$}
\newcommand{\vfifty}{V$_{\rm 50}$}
\newcommand{\vtwenty}{V$_{\rm 20}$}

\newcommand{\Rmax}{R$_{\rm max}$}
\newcommand{\RHI}{R$_{\rm HI}$}
\newcommand{\Vmax}{$V_{\rm max}$}

\def\kms{km$\,$s$^{-1}$}
\newcommand{\vflat}{V$_{\rm flat}$}

\newcommand{\hMpc}{{\ifmmode{h^{-1}{\rm Mpc}}\else{$h^{-1}$Mpc}\fi}}
\newcommand{\hkpc}{{\ifmmode{h^{-1}{\rm kpc}}\else{$h^{-1}$kpc}\fi}}
\newcommand{\hMsun}{{\ifmmode{h^{-1}{\rm {M_{\odot}}}}\else{$h^{-1}{\rm{M_{\odot}}}$}\fi}}
\newcommand{\ltsima}{$\; \buildrel < \over \sim \;$}
\newcommand{\gtsima}{$\; \buildrel > \over \sim \;$}
\newcommand{\lsim}{\lower.5ex\hbox{\ltsima}}
\newcommand{\gsim}{\lower.5ex\hbox{\gtsima}}

\def\lesssim{\mathrel{\hbox{\rlap{\hbox{\lower4pt\hbox{$\sim$}}}\hbox{$<$}}}}
\def\gtrsim{\mathrel{\hbox{\rlap{\hbox{\lower4pt\hbox{$\sim$}}}\hbox{$>$}}}}

\newcommand{\beq}{\begin{equation}}
\newcommand{\eeq}{\end{equation}}
\def\beqa{\begin{eqnarray}}
\def\eeqa{\end{eqnarray}}
\def\hMpc{$h^{-1}\,{\rm Mpc}$}
\def\hkpc{$h^{-1}\,{\rm kpc}$}

\def\Reff{R$_{\rm  eff}$}
\def\Vmax{V$_{\rm max}$}
\def\Rmax{R$_{\rm max}$}
\def\Rvir{R$_{\rm vir}$}
\def\Rlast{R$_{\rm last}$}

\def\head{
 \vbox to 0pt{\vss
                   \hbox to 0pt{\hskip 440pt\rm LA-UR-10-07069\hss}
                  \vskip 25pt}}

\title[Rotation velocities and  the velocity function of dwarf galaxies]
{A matter of measurement:  rotation velocities and  the velocity function of dwarf galaxies}
\author[Brook \& Shankar]
       {Chris B. Brook$^{1,2}$\thanks{E-mail: cbabrook@gmail.com}, Francesco Shankar$^3$\thanks{E-mail: F.Shankar@soton.ac.uk}\\
$^{1}$Departamento de F\'isica Te\'orica, Universidad Aut\'onoma de Madrid, 28049 Cantoblanco, Madrid, Spain\\
$^{2}$Astro-UAM, UAM, Unidad Asociada CSIC\\
$^{3}$School of Physics and Astronomy, University of Southampton, Southampton SO17 1BJ, UK\\
}

\begin{document}

\date{Accepted XXXX . Received XXXX; in original form XXXX}

\pagerange{\pageref{firstpage}--\pageref{lastpage}} \pubyear{2010}

\maketitle

\label{firstpage}

%\clearpage

%%%%%%%%%%%%%%%%%%%%%%%%%%%%%%%%%%%%%%%%%%%%%%%%%%%
\begin{abstract}
The  velocity function derived from large scale surveys can be compared with the predictions of $\Lambda$CDM cosmology, by matching  the measured  rotation velocities  \vrot\ of galaxies to the maximum circular velocity of dark matter (DM) halos \Vmax. For  \vrot$<$50\,kms$^{-1}$, a major discrepancy arises between the observed and $\Lambda$CDM velocity functions. However,  the manner in which  different observational measures of \vrot\  are associated with \Vmax\ is not straight forward in  dwarf galaxies. 
  We instead relate galaxies to DM halos using the  empirical  baryon-mass to halo-mass  relation, and  show that different observational measures of \vrot\   result  in very different velocity functions. We show how the W50 velocity function, i.e. using the HI profile line width at 50\% of peak HI flux to measure \vrot, can be reconciled with a \lcdm\ cosmology.    Our semi-empirical methodology allows us to determine the region of rotation curves that are  probed by HI measurements (\RHI), and shows that the \vrot\ of dwarfs are generally  measured at a fraction of \Rmax, explaining their tendency to have rising rotation curves.   We provide fitting formulae for relating \RHI\ and R$_{\rm eff}$ (the effective radius) to the virial radius of DM halos. 
   To continue to use  velocity functions as a probe of \lcdm\ cosmology, it is necessary to be  precise about how the different measures of rotation velocity are probing the mass of the DM halos, dropping the assumption that any measure of rotational velocity can be equally used as a proxy for \Vmax. 
\end{abstract}

%%%%%%%%%%%%%%%%%%%%%%%%%%%%%%%%%%%%%%%%%%%%%%%%%%%
\noindent
\begin{keywords}
 galaxies: evolution - formation - haloes cosmology: theory - dark matter
 \end{keywords}

%%%%%%%%%%%%%%%%%%%%%%%%%%%%%%%%%%%%%%%%%%%%%%%%%%%
\section{Introduction} \label{sec:introduction}
%%%%%%%%%%%%%%%%%%%%%%%%%%%%%%%%%%%%%%%%%%%%%%%%%%%
The rotation velocity \vrot\ of galaxies provides a measure of the mass within the radius at which the velocity measurement was made. The velocity function of galaxies, the number density of galaxies as a function of their velocity,   is thus a close kin to the mass function, the number density of galaxies as a function of  mass. 

The observed velocity function of galaxies has been derived by large scale HI surveys,  using the HI line width as a measure of rotational velocity.
The velocity (and mass) functions for dark matter halos in a  \lcdm\ cosmology can be readily derived for DM only simulations, with circular velocities defined as  $\sqrt{G{\rm M(r)/r}}$ where $G$ is the gravitational constant, and M(r) the mass inclosed within each radius r.  Thus, one can compare the velocity function of observed galaxies to that of  dark matter halos in simulations, by matching the observed rotational velocities  to the circular velocity of dark matter halos.  
Generally, the  maximum circular velocity, \Vmax, is chosen as representative of DM halos, although the velocity at any well defined radius could be chosen. 

 The velocity function derived using the half width of the HI line profile, measured at 50\% of the profile peak W50/2, has been shown to differ markedly from the \vmax\  function expected from \lcdm\ cosmology for galaxies with W50/2$\lsim$50\,\kms\ \citep{zavala09,zwaan10,tg11,papastergis11,klypin14}. 

However, for low mass galaxies, the relation between  W50/2,  and \Vmax\,  becomes less clear than for more massive disc galaxies for several reasons: low mass galaxies have  a higher ratio of velocity dispersion to rotation velocity, are thicker, with less regular HI discs, rising rotation curves which have not reached a maximum, and decreasing baryon fractions meaning that baryons may be confined to increasingly central  regions of the DM halo.

To make a more accurate comparison between observed and predicted velocities distributions \cite{papastergis15} relaxed the assumption that W50/2 and \vmax\ can be directly compared,   and instead matched W50/2 velocities to the velocity of DM halos at the largest radius to which the observed HI gas extends \Rlast. The mismatch between theory and observation was shown to persist, although this procedure has its own issues, such as the assumption of an NFW density profile for the DM halos, the assumption that concentration of the galaxies' halos match those of pure N-body simulations,  and the assumption that W50/2 reflects the rotational velocity of dwarf galaxies at \Rlast. 

In this study, we put aside the issue of how  to best match observed and theoretical velocities, and instead match observed galaxies to DM halos based on mass. Galaxies with regular, extended HI discs and well studied rotation curves, have been shown to follow a very tight relation between baryonic mass and circular velocity, the Baryonic Tully Fisher relation (BTFR). The small scatter in this relation provides compelling evidence that the baryonic mass of galaxies is tightly correlated with their total mass \citep[e.g.][]{mcgaugh05}. We propose that matching baryonic mass to halo mass provides a more reliable mapping between observed galaxies and DM halos, than does the mapping between velocities. Several studies have explored the relation between galaxy masses, and the mass of DM halos \citep[e.g.][]{shankar06,moster10,guo10,papastergis12}

Part of our motivation comes from the fact that the tight relation between baryon mass and rotation velocity when measured from the flat part of the rotation curve  is not well reproduced, in terms of scatter, when one looks at the HI line-widths of low mass galaxies. For example,  \cite{trachternach09} explored 11 dwarf galaxies with baryonic masses  0.14$<$\mb$<$12.92$\times$10$^{8}$\msun,  and find that 6 have well defined rotation curves and do fit on the BTFR, while 5 others have W50/2 values that fall well below the BTFR: these 5 galaxies also have poorly defined rotation curves. When matching observations to DM halos by velocities, these latter 5 galaxies  would be placed in low mass halos. 
The study of 101 dwarf galaxies in \cite{geha06}, selected as low luminosity SDSS galaxies, also shows a large scatter in the values of W20 in a sample of galaxies with small range of baryon masses. 

\begin{figure}
\hspace{-.25cm}
\includegraphics[width=.45\textwidth]{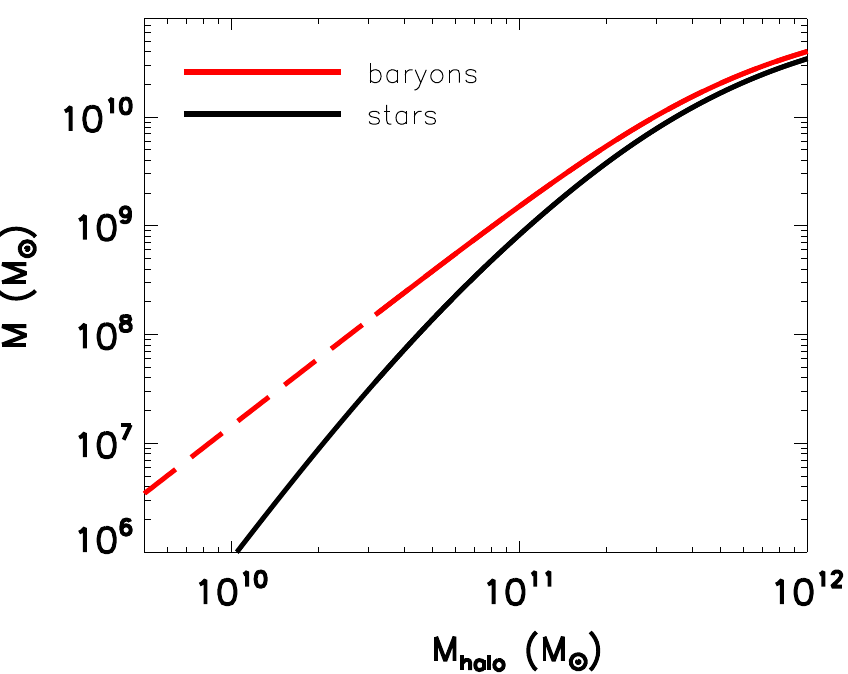}
\caption{The abundance matching relation adopted in this study, \mb-\mhalo\ (red line) and \mstar-\mhalo\ (black line). The region that is extrapolated from the Papastergis 2012  \mb-\mhalo\  relation is dashed. 
}
\label{fig:MBMH}
\end{figure}

The other point worth considering is that matching W50/2 to \vmax\  leads to many dwarf galaxies being matched to  low mass halos, implying baryon fractions that are higher than what is found in L$^*$ galaxies. For example  galaxies with W50/2 $\sim$10\,\kms, when  assuming this reflects \vmax,   are matched to halos with \mhalo$\sim$3$\times$10$^8$\msun\ (Figure 10 in \citealt{klypin14}), even though they have \mb$\sim$10$^7$\msun.

%Further, we point out that cosmological  galaxy formation simulations that match a wide range of scaling relations \citep{brook12}, match  the empirical \mb-\mhalo\ abundance matching relation  (Santos-Santos et al. submitted). 

Our contention is that the low value of W50/2 in many dwarfs is probing relatively inner regions of rising rotation curves, %This differs than studies of galaxies with regular, extended HI discs, where the BTFR  relationship holds with little scatter. 
and  therefore the \mb-\mhalo\ relation is a more robust manner for connecting galaxies to DM halos. We will see in particular that different measures of \vrot\ result in very different velocity functions, meaning that they cannot all be matched to \vmax\ equally. 
%;It is possible, of course, that the observed HI line-widths  do reflect the fact that the baryon mass-halo mass relation breaks down in low mass galaxies. 

Matching galaxies to halos using mass provides an interesting, and we believe, compelling way to compare observed and theoretical velocity functions. 
  The study is organised as follows: in  section~\ref{methods} we outline our procedure, then in  section~\ref{analysis} we show the adopted empirical relation between \mb\ and \mhalo\, and then show how \mb\ is related to \vrot\ in various observational data sets using various measures of \vrot. We then combine these relations, and show how differences in the \mb-\vrot\ relation affect the derived velocity function. We conclude that \lcdm\ is compatible with the velocity function observations, under the condition that it can explain the different BTFRs. 
In an appendix, we show that our results are robust to different data sets.

\begin{figure}
\hspace{-.25cm}
\includegraphics[width=.45\textwidth]{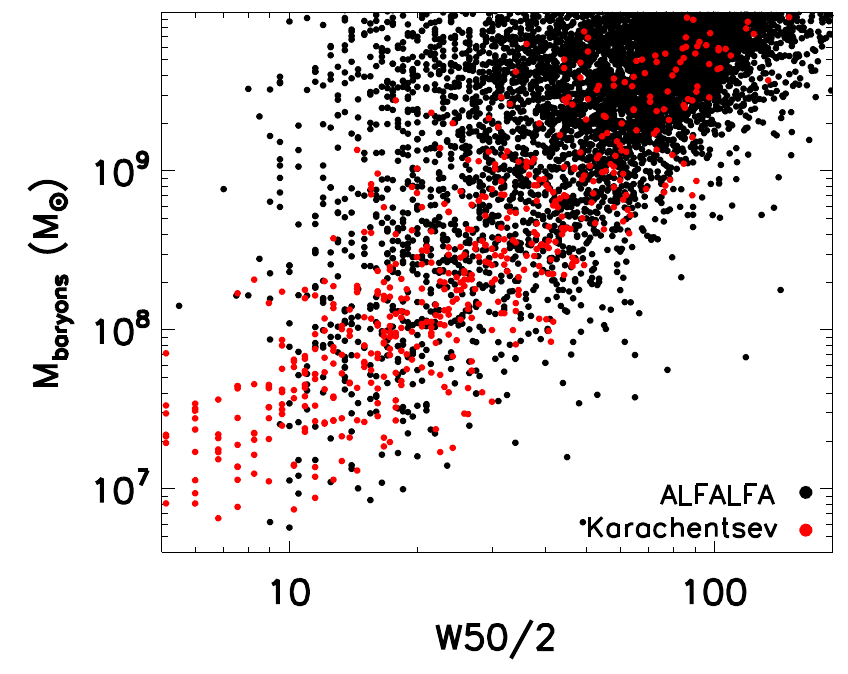}
\caption{The baryon mass versus W50/2  for ALFALFA galaxies (black dots) and the Local Volume catalogue of Karachentsev 2013 (red dots). These data sets share many galaxies. The data have similar distributions of low mass galaxies where the Local Volume statistics are better. }
\label{fig:ALFALFA}
\end{figure}

\section{Methods}\label{methods}
Our procedure for deriving the velocity function is as follows:
\vspace{.1cm}

\noindent$\bullet$ use an empirical abundance matching between \mhalo\ and \mb.\\
\noindent$\bullet$ use  observed data sets to relate \mb\ to \vrot\, where the relation is dependent on the specific manner in which rotation velocity is measured (using HI line widths or rotation curves).\\
\noindent$\bullet$ for each \mhalo\ we thus have empirically matched \mb\ and \vrot. \\
\noindent$\bullet$ we then use the mass function from N-body simulations which gives us the {\it predicted number} of halos of each mass,  and hence the predicted number of galaxies with particular values of  \mb\ and \vrot, the latter being the velocity function.
\vspace{.1cm}

\noindent In our procedure, galaxies have been matched to halos according to their mass. We have not attempted to separate satellites and central galaxies which will be explored in a subsequent study; for the theoretical mass function we simply use a power law fit to the simulations \cite{klypin11,klypin14}. With galaxies matched to halos using the adopted \mhalo\ and \mb\ relation, the resultant velocity function differs depending on the adopted relation between \mb\ to \vrot\, i.e the BTFR, which itself differs according to the manner in which \vrot\ is measured. 

For our purposes, we are interested in adopting a single relation between each of the galaxy properties \mhalo, \mstar\ and \mgas\ and then exploring the effect of convolving these relations with different BTFRs. In the appendix  we start with a different  set of relations between \mhalo, \mstar\ and \mgas\, based on the recent study of \cite{bradford15}, and then repeat our procedure. The results are very similar to those presented in the paper, indicating that our results are robust. Within the paper, we also make several other consistency checks between different data sets.  We neglect scatter in our adopted scaling relations as it will not affect our final determination of the low-velocity end of the velocity function. 

\section{Analysis}\label{analysis}
 \subsection{\mb-\mhalo}\label{mbmh}

\begin{figure}
\includegraphics[width=.45\textwidth]{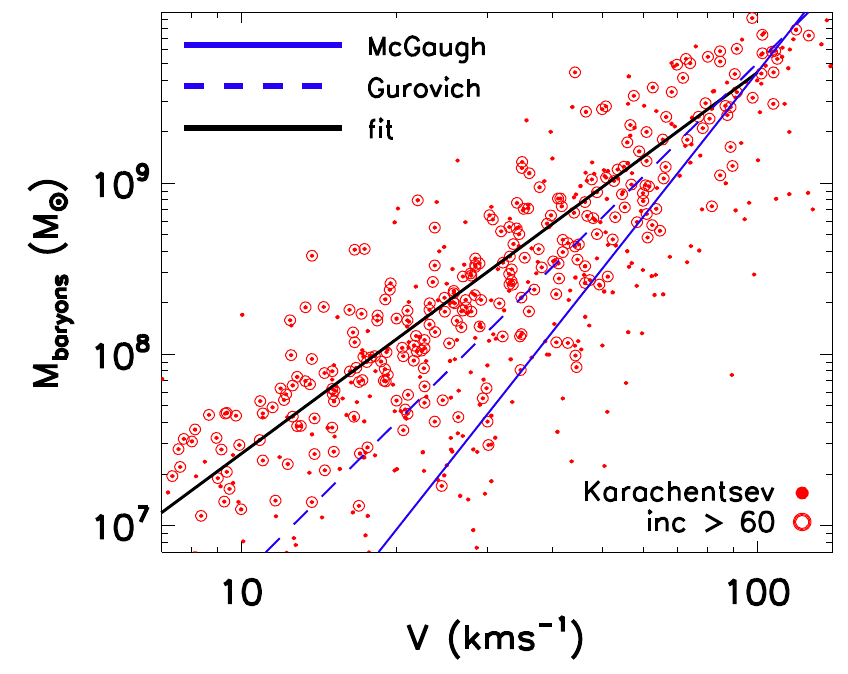}
\caption{The baryonic Tully-Fisher relation (BTFR) for the Karachentsev 2013 galaxies, with a fit to galaxies which have inclinations inc$>$60$^{\circ}$. We use \vfifty,  the W50 line half widths adjusted for inclination and velocity dispersion. 
Also shown are the BTFRs of McGaugh 2012, derived using \vflat\ and from Gurovich 2010, derived using W20.} 
\label{fig:LVbtfr}
\end{figure}

We take a fit to the \mb-\mhalo\ relation from \cite{papastergis12}, using the equation:

\begin{equation}
 {\rm M}_{\rm b} =0.0635 {\rm M}_{\rm halo} \left[\left(\frac{{\rm M}_{\rm halo} }{10^{11.65}}\right)^{-1.057} + \left(\frac{{\rm M}_{\rm halo} }{10^{11.65}}\right)^{0.556}\right]^{-1}
\label{eq:mbmh}
\end{equation} 

\noindent This has the same functional form as the \mstar-\mhalo\ relation as \cite{moster10}. The relation is shown as a red line in Figure~1, with the region that is extrapolated from the \cite{papastergis12} relation shown as a dashed line. 
We then compute the gas and stellar composition of the baryons using equation 1 of Papastergis 2012, 
\begin{equation}
{\rm log}_{\rm 10}({\rm M}_{\rm HI}/{\rm M}_{\rm *})=-0.43{\rm log}_{10}({\rm M}_{\rm *})+3.75 
\label{eq:hifrac}
\end{equation}

\noindent The adopted relation between gas and stellar mass of galaxies is also consistent with the empirical relation derived  by \cite{peeples11}.
The residual \mstar-\mhalo\ relation, shown as a black line in Figure~1, is  close to the \cite{guo10} relation. It has been shown that the extrapolation of \cite{guo10} to lower mass galaxies gives a good approximation of abundance matching results for Local Group galaxies \citep{brook14}.

Galaxies are matched down to halo masses of around 7$\times$10$^9$. Below this limit, several studies have shown that galaxy formation is suppressed by the early re-ionisation   of the intergalactic medium \citep{bullock00,somerville02,benson02}. 
% This  fits what is seen in the Local Group and the satellites of the MW and M31 (see e.g. \citealt{Brook14,Brook15}.

\begin{figure}
\hspace{0.cm}\includegraphics[width=.45\textwidth]{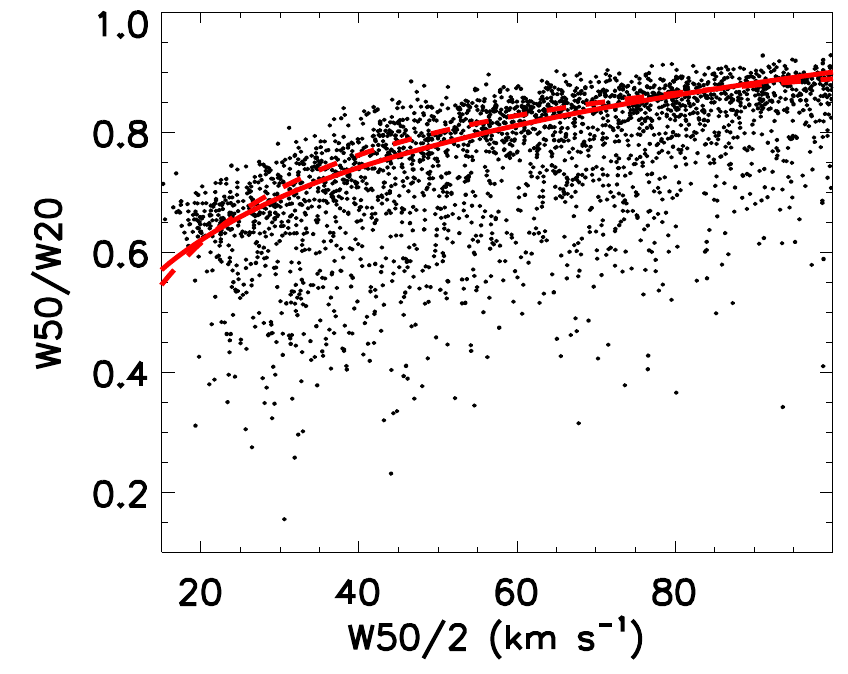}
\caption{HIPASS data showing the difference between HI line width measured at 20\% and 50\% of the peak flux. We plot W50/2 versus W50/W20.
 For low mass galaxies, there is a  big difference in the two indicators of rotation velocity, with a large tail toward low values of W50/W20. The fitted red line will be used to show the change in the velocity function when using W20, as compared to W50, as an indicator of rotation velocity. The red dashed line shows the result of using W20 = W50+25\,\kms. }
  \label{fig:w50w20}
\end{figure}

\subsection{\mb-\vrot: measures of the BTFR}
We first explore the relation between W50/2 and baryonic mass for ALFALFA galaxies \citep{haynes11}, and for the Local Volume (LV) galaxies as compiled by \cite{karachentsev13}. We calculate \mstar\ for the ALFALFA galaxies following \cite{bell03} and calculate HI mass \mHI\ =2.36e5d$^2$$\times$flux, where flux is the HI flux and d is the distance in Mpc, then \mb=\mstar+4/3\mHI. 
 We plot, in Figure~2,  \mb\ versus W50/2, where the ALFALFA galaxies are shown as black circles and the LV galaxies as red circles. The distributions are similar for the low velocity region we are interested in, and in fact there is significant overlap in the data sets. 
 
 We define 
  \begin{equation}
 {\rm V}_{50}=\sqrt{({\rm W50}/2/sin({\rm inc}))^2-8^2}
 \label{incdisp}
 \end{equation}  
 \noindent taking account of inclination and velocity dispersion in deriving a measure of rotation velocity.  
 In Figure~3 we then take the LV galaxies  and plot the \vfifty\ BTFR (red dots) and use  galaxies that have inclinations inc$>$60$^\circ$ (red circles) to  make a fit (black line):
 \begin{equation}
 {\rm log}_{10}{\rm M}_{\rm b}=5.19+2.22{\rm log}_{10}{\rm V}_{50}
 \label{eq:w50btfr}
\end{equation}

Also shown in Figure~3 are two BTFRs taken from the literature;  the BTFR derived using the flat part of the rotation curve \vflat\ of galaxies with regular, extended HI discs and well resolved rotation curves,  ${\rm log}_{10}{\rm M}_{\rm b}=2.01+3.82{\rm log}_{10}{\rm V}_{\rm flat}$ \citep{mcgaugh12}; and the BTFR derived using inclination and dispersion adjusted W20 measurements,    ${\rm log}_{10}{\rm M}_{\rm b}=3.7+3{\rm log}_{10}{\rm V}_{\rm 20}$  \citep{gurovich10}. 
% It is established that \vtwenty\ gives a flatter BTFR than \vflat\ \citep[e.g.][]{mcgaugh12}.  
%The BTFR we derived using \vfifty\ of the larger sample of galaxies is even flatter.  

\subsection{W50 compared to W20}
%The velocity function derived by \cite{zwaan10} using HIPASS data is very similar to that using ALFALFA data \citep{papastergis15}: the data sets are compatible. 
Here, we  highlight the difference between W50 and W20, in low mass galaxies, i.e. the difference between measuring the HI line-widths at 20\% and 50\% of the peak value. We use data  taken from the HOPCAT catalogue \citep{doyle05} of the HIPASS survey, but also show the relation from the recent study of \cite{bradford15}, using the ALFALFA catalogue. 

Figure~\ref{fig:w50w20} shows the ratio W50/W20 as a function of W50/2.  For low mass galaxies, there is a  significant difference in the two indicators of rotation velocity, with a large tail toward low values of W50/W20. We will use the fitted red line,
\begin{equation}
 {\rm W50/W20} = 0.1 +   0.4{\rm log}_{10}({\rm W50}/2) 
 \label{eq:w50w20}
\end{equation}
 \noindent to adjust our velocity function from one based on W50, to one based on W20, shown as dashed red lines  in figures 4, 6 \& 7, highlighting the effect of choosing W50 over W20 as an indicator of rotation velocity. This adjustment is almost identical to simply using W20=W50+25\kms (dashed red line in Figure~\ref{fig:w50w20}), which is the typical difference between these measures of \vrot\ according to both \cite{bradford15} and \cite{koribalski04}. 

%In Figure~\ref{fig:hopcat} we plot \vfifty\ (black dots) and \vtwenty\ (red dots) versus \mb\ for the HOPCAT data, including only galaxies with inc$>$60$^\circ$. where \vtwenty\ is adjusted for inclination and dispersion in the same way as \vfifty\, i.e. following equation~\ref{incdisp}. The black line is not a fit to the data, but is the adopted W50 BTFR fit from Figure~\ref{fig:LVbtfr}, i.e. equation~\ref{eq:w50btfr}. The red dashed line shows the adjustment from W50 to W20, coming from equation~\ref{eq:w50w20}. The different data sets are comparable.

%\begin{figure}
%\includegraphics[width=.45\textwidth]{hopcatBTF.pdf}
%\caption{HIPASS data taken from the HOPCAT catalogue, showing galaxies with inc$>60^{\circ}$. The BTFR is shown as black and red dots for \vfifty\ and \vtwenty. The lines are not  fits to  the data: the black line shows the fit to the Karachentsev 13 data, as shown in Figure 2, while the dashed line is the W50 to W20 adjustment to the black line. }
%\label{fig:hopcat}
%\end{figure}

\begin{figure}
\includegraphics[width=.45\textwidth]{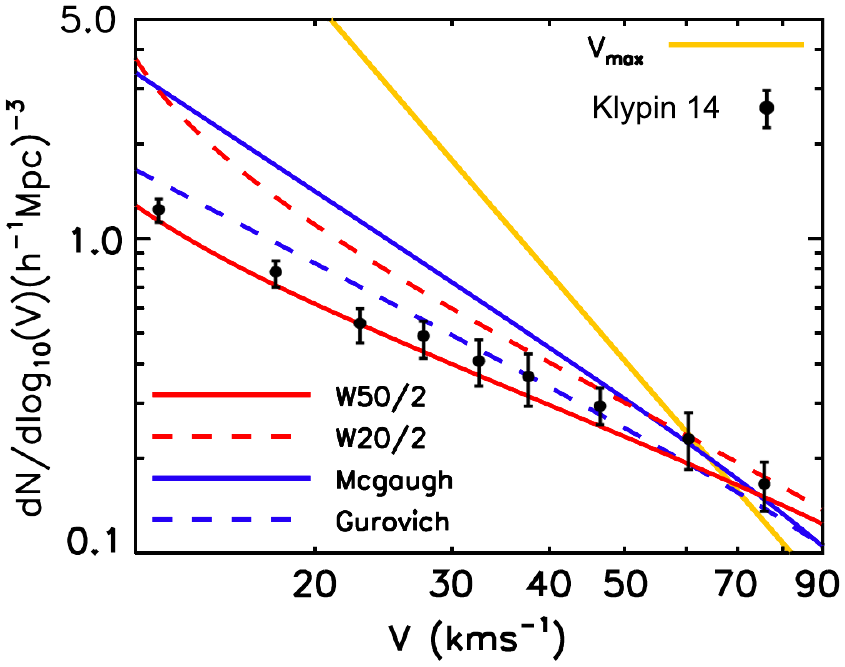}
\caption{The yellow line shows the V$_{\rm max}$ velocity function of DM only simulations. We then show the velocity function predicted by $\Lambda$CDM cosmology when making different assumptions about the BTFR, having matched observed baryon masses to DM halo masses. Using the W50/2 BTFR, equation~\ref{eq:w50btfr}, is shown as the red line. The W20/2 BTFR, derived using equation~\ref{eq:w50w20} is shown as a red dashed line. Results of assuming the BTFRs of Mcgaugh (2012) and Gurovich et al. (2010) are shown as blue solid and blue dashed lines. The observational results from Klypin et al. (2014) are shown as black circles with error bars}
\label{fig:VFn}
\end{figure}

\subsection{The Velocity Function}
In Figure~\ref{fig:VFn}, we explore the effect of different BTFRs on the velocity function, when the BTFRs are used with our adopted \mb-\mhalo\ relation. In each case  \vrot\ values are adjusted to line of sight velocities V$_{\rm los}$ by averaging over all inclinations to allow  direct comparison with the observed W50/2 velocity function\footnote{We have not included a dispersion correction as the observed velocity functions use W50/2 directly. We derived the \vfifty\ BTFR having used an inclination and  dispersion corrected W50/2, because  such relation follows a single power law, with lower scatter, than the W50/2 BTFR.  The re-inclusion of the velocity dispersion when deriving the velocity function is what  causes the upturn in the W50/2 or W20/2 velocity functions at low velocities. The final result will not change if the velocity dispersion is maintained throughout, so long as one selects an appropriate (but more complicated) relation between baryon mass and measured W50/2.}\citep{klypin14,papastergis15}.

The yellow line in Figure~\ref{fig:VFn},   shows the \vmax\ velocity function of DM only simulations, while the the velocity function derived from the W50/2 BTFR, equation~\ref{eq:w50btfr}, is shown as the red line. The velocity function derived from the W20/2 BTFR using equations~\ref{eq:w50btfr}~\&~\ref{eq:w50w20} is shown as a red dashed line. Results of assuming the BTFRs of \cite{mcgaugh12}, and \cite{gurovich10} are shown as blue solid and blue dashed lines respectively. The effect of different BTFRs on the velocity function is significant.

Consider the difference between the velocity function derived using the \vflat\ BTFR, i.e. the blue solid line, as compared to the yellow line which shows the \vmax\ velocity function predicted from pure N-body simulations.  The difference is large,  yet hydrodynamical galaxy formation simulations within a \lcdm\ cosmological context have been shown to simultaneously match both the \mb\-\mhalo\ {\it and} the  \vflat\ BTFR  relation (Santos-Santos et al. submitted), which is sufficient to reconcile the lines. 

%Therefore, when considering galaxy formation processes, the predictions of \lcdm\ are not embodied by the yellow line that is derived from purely DM only simulations: simulated galaxies within a \lcdm\ context are entirely compatible with the \vflat\ velocity function. 

%It has been shown that simulated galaxies match the \vflat\ BTFR (\citealt{brook12,aumer13}; Santos-Santos et al. 2015). The latest  of these studies also shows that the scatter in the simulations around the \vflat\ BTFR is small, and that the simulations also match the empirical \mb-\mhalo\ relation adopted in this current study. 

\begin{figure}
\hspace{0cm}\includegraphics[width=.45\textwidth]{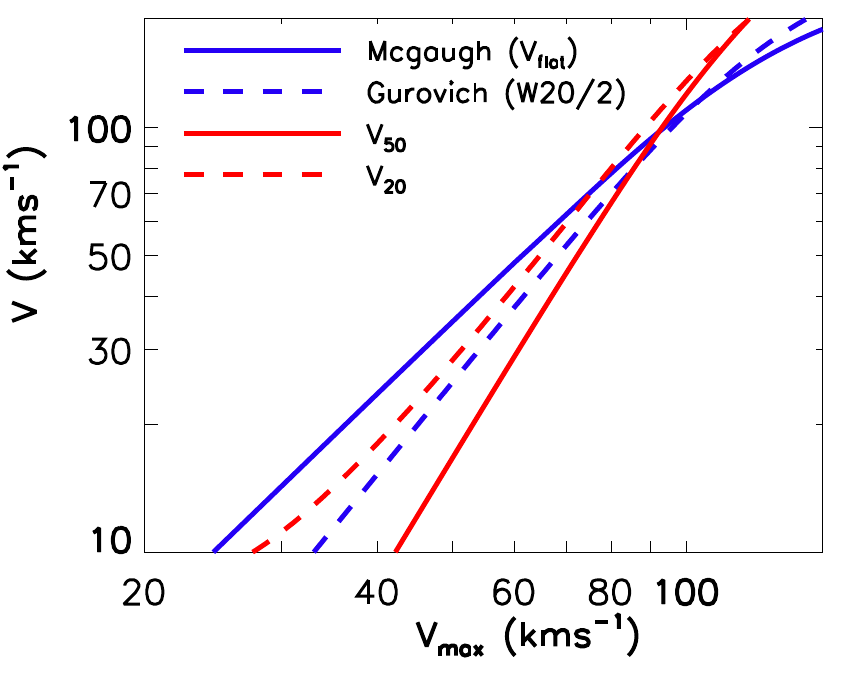}
\caption{
We plot \vrot\ as a function of the \vmax\ of the matched halo, under different assumptions for the BTFR after matching galaxies to halos on the basis of mass. We use 2 different empirical BTFRs from the literature,  shown as blue lines. We also use the fit to W$_{50}$ from figure 2  (red line) and the adjustment from W50 to W20 as shown in Figure 4 (red dashed line). 
 }\label{fig:vrvmax}
\end{figure}

The red line, derived using the \vfifty\ BTFR,  shows the W50/2 velocity function, which is very similar to the observed W50/2 velocity function \citep[e.g.][shown as black cycles with error bars in Figure~\ref{fig:VFn}]{klypin14}. 
%Yet, if one can reproduce the W50/2 BTFR within \lcdm\ cosmology, then one will reconcile the different velocity functions. %This is likely  gas dispersions and velocities in populations of low mass galaxies.

%one needs to show that the galaxy formation models can also match the W50/2 BTFR relation, as well as the \vflat\ BTFR.  
%, on top of their demonstrated ability to match the \vflat\ BTFR for galaxies with extended HI discs and well studies rotation curves.

%At first, this may appear unsurprising considering that the observed velocity function was derived using the same  W50/2 data. 
%But the {\it number} of galaxies at each velocity in the red line has been taken from \lcdm\ cosmology. 
% For this W50/2 velocity function to be compatible with \lcdm\, one would need a similar relation difference the W50/2 and \vflat\ BTFRs of simulated, low mass galaxies, as is found in observed galaxies. 

\subsection{Relating various measures of V$_{\rm rot}$ to V$_{\rm max}$}
Having matched galaxies to DM halos based on mass, we can explore the implied links between various measures of \vrot\ for the observed galaxies, and the \vmax\ of matched DM halos. This is shown in Figure~\ref{fig:vrvmax} which shows the results for the different BTFRs. The \vrot\ taken from the two empirical BTFRs, as seen in earlier plots, are shown  as a function of the \vmax\ of the matched DM halos (blue lines). Results using the \vfifty\ BTFR, i.e. equation~\ref{eq:w50btfr}, are shown as the solid red line, and the adjustment from W50 to W20, equation~\ref{eq:w50w20} as the red dashed line.

%The  red line is similar to the relation found in \cite{papastergis15}, who matched observed W50  velocities to the circular velocity of DM halos measured at \Rlast\ of the observed galaxies. In deriving this relation, we used a different process than \cite{papastergis15} who matched velocities directly, while we matched by mass before applying the W50/2 BTFR. The consistency  means that the W50/2 BTFR that we adopted, i.e. equation~\ref{eq:w50btfr} also works for the data set as  selected by \cite{papastergis15}, which is another important self-consistency test for our study.
%, as the flat nature of the W50/2 BTFR as compared to the \vflat\ BTFR is really at the heart of this study.

%The blue line in Figure~\ref{eq:fig:vrvmax} shows the relationship between \vflat\ and \vmax\ for galaxies that follow the BTFR of \cite{mcgaugh12}, i.e. the relation for galaxies with regular . 

\begin{figure}
\hspace{0cm}\includegraphics[width=.45\textwidth]{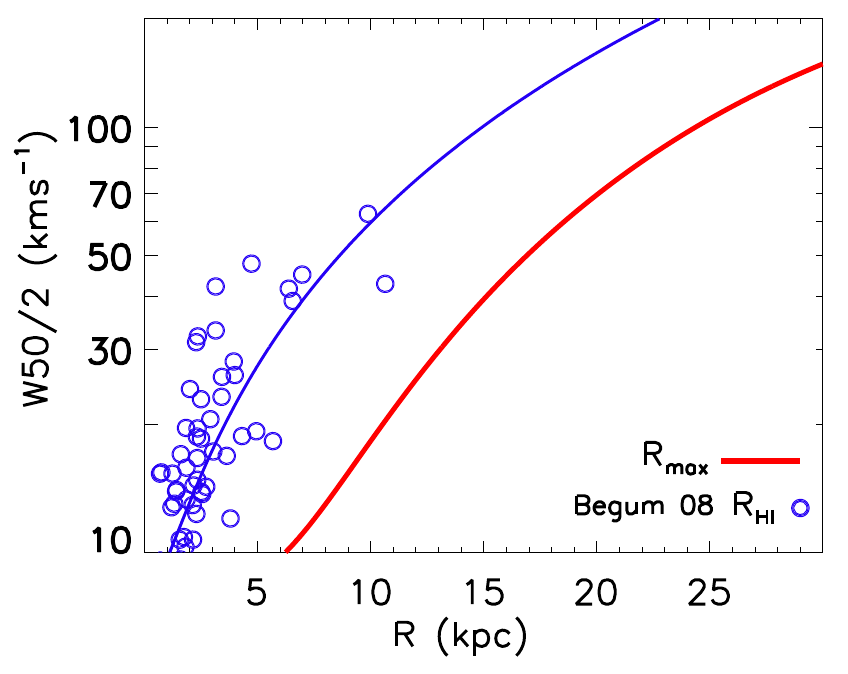}
\caption{The red line is the \Rmax\ of  DM halos, plotted against the value of W50/2 for the matched galaxies. 
The blue circles show the radial extent of the HI discs (R$_{\rm HI}$) versus W50/2 of the FIGGs (Begum et al. 2008) sample.  The blue line takes the fit of R$_{\rm HI}$ versus HI gas mass from Begum (2008), and converts from HI mass to W50/2 using equations~\ref{eq:hifrac}~\&~\ref{eq:w50btfr}.}\label{fig:rmax}
\end{figure}

\subsection{The extent of low mass galaxies}
Finally, we highlight the difference between  the extent of HI gas in galaxies, and the \Rmax\ of their matched DM halos, where \Rmax\ is the radius at which \vmax\ is reached. In Figure~\ref{fig:rmax},  W50/2 is plotted  as a function of R$_{\rm HI}$, defined as the radius at which the HI gas density reached 19 atoms cm$^{-2}$, for the FIGGs sample of galaxies \citep{begum08} shown as blue circles, a representative subset of the compiled \cite{karachentsev13}
data.

%The red line shows the result of using  the W50/2  via the BTFR in equation~\ref{eq:w50btfr}, and   R$_{max}$ is the radius at which  the matched DM halo reaches \vmax. 
%The blue circles show the radial extent of the HI discs (R$_{\rm HI}$) versus W50/2 of the FIGGs (Begum et al. 2008) sample.  The blue line shows takes the fit of R$_{\rm HI}$ versus HI gas mass from equation 1 in Begum (2008), and converts from HI mass to W50/2 using the  BTFR from equation~\ref{eq:w50btfr} and HI mass fraction from equation~\ref{eq:hifrac} of this paper: the fact that the circles follow the blue line  shows the internal self consistency between the data sets, i.e. that the Begum 08 data is consistent with the LV data in terms of the baryon mass-W50/2 relation.

The red line in  Figure~\ref{fig:rmax} shows the \Rmax\ of  DM halos, plotted against the value of W50/2 for the matched galaxies. The blue circles show the radial extent of the HI discs (R$_{\rm HI}$) versus W50/2 for the FIGGs  sample.
%Galaxies do not extend to \Rmax, assuming that they can be matched to halos by mass. 

%In massive disc galaxies, baryon fractions are high: as baryons dissipate and become centrally concentrated, rotation curves are steep, and so the \vrot\ of galaxies  is reasonably reflective of \vmax\ of the matched DM halo, once baryons have been modelled. But in low mass galaxies, baryon fractions are low meaning and baryons are merely tracers of halo mass: in his case, measured values of \vrot\ are generally lower than \vmax, often significantly lower. This is why low mass galaxies generally have rising rotation curves at their last measured points. 

The blue line  in  Figure~\ref{fig:rmax}  uses  equation~1 of \cite{begum08} which relates HI mass to  R$_{\rm HI}$, along with our equations~\ref{eq:hifrac}~\&~\ref{eq:w50btfr}. This is a good consistency check for the data sets.
%: assuming that the Local Volume galaxies have a similar relation between HI mass and  R$_{\rm HI}$ as the FIGGS data set, then the two must have a similar relation between W50/2 and \mb\ for blue line to follow the same trend as the circles.

\begin{figure}
\hspace{0cm}\includegraphics[width=.45\textwidth]{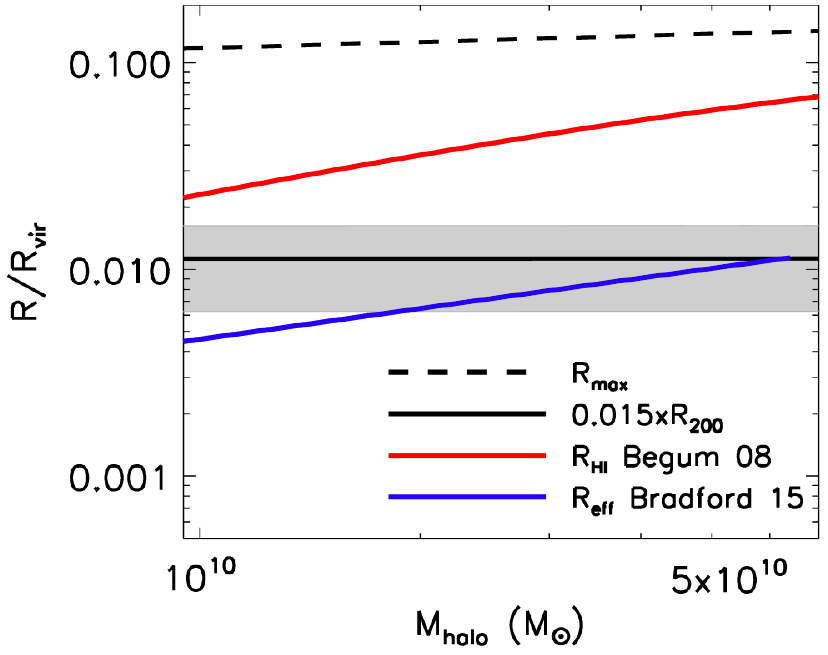}
\caption{The ratios \RHI/\Rvir  and \Reff/\Rvir\ for observed galaxies and their matched halos. \RHI\ comes from Begum (2008) while \Reff\ from Bradford et al. 2015. The black line  and grey area at 0.01$\mp$0.005$\times$ R$_{200}$, is the  relation found by  Kravtsov (2013) for \Reff\ of more massive galaxies. }
\label{RRvir}
\end{figure}

In Figure~\ref{RRvir}, we plot the ratios  \RHI/\Rvir  and \Reff/\Rvir\ for observed galaxies and their matched halos. \RHI\ comes from the FIGGS sample Begum (2008) while \Reff\ from the sample of  Bradford et al. 2015. The black line  and grey area at 0.015$\mp$0.005$\times$ R$_{200}$, is the  relation found by  \cite{kravtsov13} for \Reff\ of more massive galaxies. A constant ratio of \Reff/\Rvir\  does not hold for lower mass galaxies, which are  confined to more central regions within their host halos. 
The ratios are well fit power laws
\begin{equation}
 {\rm log}_{10}\rm{R}_{\rm HI}/{\rm R}_{\rm vir} = -7.65+0.6 log_{10} {\rm M}_{\rm halo}
 \end{equation}
 
 and
\begin{equation}
 {\rm log}_{10}{\rm R}_{\rm eff}/{\rm R}_{\rm vir}= -7.25+0.49 {\rm log}_{10} {\rm M}_{\rm halo}
 \end{equation}

\section{Discussion}
When matching W50/2 velocities of observed  galaxies to \vmax\ of DM halos, one finds a significant excess in the  {\it number} of DM halos with \vmax$\lsim$\,50\kms. 
This mismatch  occurs where  halos are too massive to have had galaxy formation prevented by re-ionisation; from this perspective the observed velocity function indicates that there is a ``too big to fail" problem  \citep{zavala09,klypin14,papastergis15}.
 
In this study we instead match  observed galaxies to DM halos based on baryonic mass, and explored the impact of the various forms of the BTFR  on the velocity function that is expected in \lcdm\ cosmology.  The  BTFR  derived from W50/2 values taken from large surveys of galaxies,  results in a very different prediction for the velocity function than the one  inferred from the BTFR  derived using the \vflat\ values of dwarf galaxies that have extended, regular HI rotation fields.  The question of whether the observed  velocity function is consistent with \lcdm\ cosmology  may be  reduced to the question of whether the various forms of the BTFR can be reproduced within a \lcdm\ cosmology. 

Yet it is has already been shown that  hydrodynamical  cosmological galaxy formation simulations are able to simultaneously match  the \vflat\ BTFR and  \mb-\mhalo\  relation (Santos-Santos et al. 2015 submitted, see also \citealt{brook12,aumer13}), which is sufficient to reconcile the observed  \vflat\ velocity function with \lcdm\ predictions. Therefore, when considering galaxy formation processes, the predictions of \lcdm\ are not embodied by the yellow line that is derived from purely DM only simulations: simulated galaxies within a \lcdm\ context are  compatible with the observed \vflat\ velocity function.  

Whether  the W50/2 velocity function is compatible with \lcdm\ cosmology becomes a matter of understanding  how measures of W50/2 in a large sample of galaxies relate to measures of \vflat\ in galaxies with extended HI discs and regular rotation curves.%: a matter of astrophysics rather than cosmology 

When matching galaxies to DM halos by mass,  the problem lies with  understanding observational measurements of \vrot\,  rather than  a shortfall in the number of galaxies that are ``too big to fail". This may  require that the halos have DM cores, rather than cusps. Yet we already have significant evidence that many dwarf disc galaxies having slowly rising rotation curves, possibly indicative of cored DM profiles. 

Various proposals for explaining the existence of cores have been proposed, both within the \lcdm\ paradigm \citep[e.g.][]{navarro96b,read06,governato10,pontzen12}, or by taking an alternative form of DM \citep[e.g.][]{zavala09,vogelsberger12,schneider14}. In an accompanying paper (Brook \& Di Cintio 2015 submitted), we will compare how  galaxy cusps and cores are reflected in the velocity function, and then proceed to highlight how different core formation mechanisms may leave signatures in the detailed shape of the velocity function.

%Massive disc galaxies have large baryon fractions, and baryons dissipate to the centre of DM halos: this means that the rotation curve rises sharply, and the observed \vrot\ can provide a reasonable proxy for \vmax. Regular rotation within HI discs, with relatively small  velocity dispersion, mean that the various measures of \vrot\ give similar results for massive disc galaxies.
According to our analysis, \vrot\ is not measured at R$_{max}$ in dwarf galaxies. 
 We provide relations between \RHI\ and \Reff\ of observed galaxies, and \Rvir\ of their matched DM halos, and showed that dwarfs are confined to more central regions of their matched DM halos than  more massive galaxies. This explains why rotation curves of dwarf galaxies are generally  rising at the last measured point.
As one does not reach \Rmax\, the measured \vrot\ is dependent on the extent of the HI gas, as well as its detailed structure.  Scatter begins to emerge in the various BTFRs for low mass galaxies, and the different measures of \vrot\ become increasingly different, leading to different velocity functions. 
%More detailed understanding of HI gas, its three dimensional morphology and velocities, is required before we can proceed with using HI line-widths of dwarf galaxies as cosmological probes. 

We performed several consistency checks between different data sets from the literature. In the appendix, we have taken this further,  redoing our analysis using the recent data from \cite{bradford15}, who measured the HI line-widths of a large sample of dwarf galaxies selected from the SDSS catalog. The results, and in particular the large effect of different measures of \vrot\ on the velocity function, are unchanged from the analysis in the paper. 

Our study highlights that if the velocity function is  to be used as a probe of cosmological paradigm, we need to be precise in relating measures of \vrot\ with the \vmax\ of DM halos. Not all measures of \vrot\ are equal and not all  BTFRs have the same slope, particularly in low max galaxies. 
In a forthcoming paper, we will explore various forms of the BTFR within hydrodynamical, cosmological simulations, looking at how the different measures of the BTFR can be reconciled within a \lcdm\ context.

\section*{Acknowledgements}
CB thanks the MICINN (Spain) for the financial support through the MINECO grant AYA2012-31101 and  the Ramon y Cajal program. He further thanks the DARK cosmology centre for their kind hospitality. 
%%%%%%%%%%%%%%%%%%%%%%%%%%%%%%%%%%%%%%%%%%%%%%%%%%%
\bibliographystyle{mn2e}
\bibliography{archive}

\begin{thebibliography}{37}
\expandafter\ifx\csname natexlab\endcsname\relax\def\natexlab#1{#1}\fi

\bibitem[{{Aumer} {et~al}\mbox{.}(2013){Aumer}, {White}, {Naab}, \&
  {Scannapieco}}]{aumer13}
{Aumer} M., {White} S.~D.~M., {Naab} T., {Scannapieco} C., 2013, \mnras, 434,
  3142

\bibitem[{{Begum} {et~al}\mbox{.}(2008){Begum}, {Chengalur}, {Karachentsev},
  {Sharina}, \& {Kaisin}}]{begum08}
{Begum} A., {Chengalur} J.~N., {Karachentsev} I.~D., {Sharina} M.~E., {Kaisin}
  S.~S., 2008, \mnras, 386, 1667

\bibitem[{{Bell} {et~al}\mbox{.}(2003){Bell}, {McIntosh}, {Katz}, \&
  {Weinberg}}]{bell03}
{Bell} E.~F., {McIntosh} D.~H., {Katz} N., {Weinberg} M.~D., 2003, ApJL, 585,
  L117

\bibitem[{{Benson} {et~al}\mbox{.}(2002){Benson}, {Frenk}, {Lacey}, {Baugh}, \&
  {Cole}}]{benson02}
{Benson} A.~J., {Frenk} C.~S., {Lacey} C.~G., {Baugh} C.~M., {Cole} S., 2002,
  \mnras, 333, 177

\bibitem[{{Bradford} {et~al}\mbox{.}(2015){Bradford}, {Geha}, \&
  {Blanton}}]{bradford15}
{Bradford} J.~D., {Geha} M.~C., {Blanton} M.~R., 2015, ArXiv e-prints

\bibitem[{{Brook} {et~al}\mbox{.}(2014){Brook}, {Di Cintio}, {Knebe},
  {Gottl{\"o}ber}, {Hoffman}, {Yepes}, \& {Garrison-Kimmel}}]{brook14}
{Brook} C.~B., {Di Cintio} A., {Knebe} A., {Gottl{\"o}ber} S., {Hoffman} Y.,
  {Yepes} G., {Garrison-Kimmel} S., 2014, \apjl, 784, L14

\bibitem[{{Brook} {et~al}\mbox{.}(2012){Brook}, {Stinson}, {Gibson}, {Ro{\v
  s}kar}, {Wadsley}, \& {Quinn}}]{brook12}
{Brook} C.~B., {Stinson} G., {Gibson} B.~K., {Ro{\v s}kar} R., {Wadsley} J.,
  {Quinn} T., 2012, MNRAS, 419, 771

\bibitem[{{Bullock} {et~al}\mbox{.}(2000){Bullock}, {Kravtsov}, \&
  {Weinberg}}]{bullock00}
{Bullock} J.~S., {Kravtsov} A.~V., {Weinberg} D.~H., 2000, \apj, 539, 517

\bibitem[{{Doyle} {et~al}\mbox{.}(2005){Doyle}, {Drinkwater}, {Rohde},
  {Pimbblet}, {Read}, {Meyer}, {Zwaan}, {Ryan-Weber}, {Stevens}, {Koribalski},
  {Webster}, {Staveley-Smith}, {Barnes}, {Howlett}, {Kilborn}, {Waugh},
  {Pierce}, {Bhathal}, {de Blok}, {Disney}, {Ekers}, {Freeman}, {Garcia},
  {Gibson}, {Harnett}, {Henning}, {Jerjen}, {Kesteven}, {Knezek}, {Mader},
  {Marquarding}, {Minchin}, {O'Brien}, {Oosterloo}, {Price}, {Putman}, {Ryder},
  {Sadler}, {Stewart}, {Stootman}, \& {Wright}}]{doyle05}
{Doyle} M.~T. {et~al.}, 2005, \mnras, 361, 34

\bibitem[{{Geha} {et~al}\mbox{.}(2006){Geha}, {Blanton}, {Masjedi}, \&
  {West}}]{geha06}
{Geha} M., {Blanton} M.~R., {Masjedi} M., {West} A.~A., 2006, \apj, 653, 240

\bibitem[{{Governato} {et~al}\mbox{.}(2010){Governato}, {Brook}, {Mayer},
  {Brooks}, {Rhee}, {Wadsley}, {Jonsson}, {Willman}, {Stinson}, {Quinn}, \&
  {Madau}}]{governato10}
{Governato} F. {et~al.}, 2010, Nature, 463, 203

\bibitem[{{Guo} {et~al}\mbox{.}(2010){Guo}, {White}, {Li}, \&
  {Boylan-Kolchin}}]{guo10}
{Guo} Q., {White} S., {Li} C., {Boylan-Kolchin} M., 2010, MNRAS, 404, 1111

\bibitem[{{Gurovich} {et~al}\mbox{.}(2010){Gurovich}, {Freeman}, {Jerjen},
  {Staveley-Smith}, \& {Puerari}}]{gurovich10}
{Gurovich} S., {Freeman} K., {Jerjen} H., {Staveley-Smith} L., {Puerari} I.,
  2010, \aj, 140, 663

\bibitem[{{Haynes} {et~al}\mbox{.}(2011){Haynes}, {Giovanelli}, {Martin},
  {Hess}, {Saintonge}, {Adams}, {Hallenbeck}, {Hoffman}, {Huang}, {Kent},
  {Koopmann}, {Papastergis}, {Stierwalt}, {Balonek}, {Craig}, {Higdon},
  {Kornreich}, {Miller}, {O'Donoghue}, {Olowin}, {Rosenberg}, {Spekkens},
  {Troischt}, \& {Wilcots}}]{haynes11}
{Haynes} M.~P. {et~al.}, 2011, \aj, 142, 170

\bibitem[{{Karachentsev} {et~al}\mbox{.}(2013){Karachentsev}, {Makarov}, \&
  {Kaisina}}]{karachentsev13}
{Karachentsev} I.~D., {Makarov} D.~I., {Kaisina} E.~I., 2013, \aj, 145, 101

\bibitem[{{Klypin} {et~al}\mbox{.}(2014){Klypin}, {Karachentsev}, {Makarov}, \&
  {Nasonova}}]{klypin14}
{Klypin} A., {Karachentsev} I., {Makarov} D., {Nasonova} O., 2014, ArXiv
  e-prints

\bibitem[{{Klypin} {et~al}\mbox{.}(2011){Klypin}, {Trujillo-Gomez}, \&
  {Primack}}]{klypin11}
{Klypin} A.~A., {Trujillo-Gomez} S., {Primack} J., 2011, \apj, 740, 102

\bibitem[{{Koribalski} {et~al}\mbox{.}(2004){Koribalski}, {Staveley-Smith},
  {Kilborn}, {Ryder}, {Kraan-Korteweg}, {Ryan-Weber}, {Ekers}, {Jerjen},
  {Henning}, {Putman}, {Zwaan}, {de Blok}, {Calabretta}, {Disney}, {Minchin},
  {Bhathal}, {Boyce}, {Drinkwater}, {Freeman}, {Gibson}, {Green}, {Haynes},
  {Juraszek}, {Kesteven}, {Knezek}, {Mader}, {Marquarding}, {Meyer}, {Mould},
  {Oosterloo}, {O'Brien}, {Price}, {Sadler}, {Schr{\"o}der}, {Stewart},
  {Stootman}, {Waugh}, {Warren}, {Webster}, \& {Wright}}]{koribalski04}
{Koribalski} B.~S. {et~al.}, 2004, \aj, 128, 16

\bibitem[{{Kravtsov}(2013)}]{kravtsov13}
{Kravtsov} A.~V., 2013, \apjl, 764, L31

\bibitem[{{McGaugh}(2005)}]{mcgaugh05}
{McGaugh} S.~S., 2005, \apj, 632, 859

\bibitem[{{McGaugh}(2012)}]{mcgaugh12}
{McGaugh} S.~S., 2012, \aj, 143, 40

\bibitem[{{Moster} {et~al}\mbox{.}(2010){Moster}, {Somerville}, {Maulbetsch},
  {van den Bosch}, {Macci{\`o}}, {Naab}, \& {Oser}}]{moster10}
{Moster} B.~P., {Somerville} R.~S., {Maulbetsch} C., {van den Bosch} F.~C.,
  {Macci{\`o}} A.~V., {Naab} T., {Oser} L., 2010, ApJ, 710, 903

\bibitem[{{Navarro} {et~al}\mbox{.}(1996){Navarro}, {Eke}, \&
  {Frenk}}]{navarro96b}
{Navarro} J.~F., {Eke} V.~R., {Frenk} C.~S., 1996, \mnras, 283, L72

\bibitem[{{Papastergis} {et~al}\mbox{.}(2012){Papastergis}, {Cattaneo},
  {Huang}, {Giovanelli}, \& {Haynes}}]{papastergis12}
{Papastergis} E., {Cattaneo} A., {Huang} S., {Giovanelli} R., {Haynes} M.~P.,
  2012, \apj, 759, 138

\bibitem[{{Papastergis} {et~al}\mbox{.}(2015){Papastergis}, {Giovanelli},
  {Haynes}, \& {Shankar}}]{papastergis15}
{Papastergis} E., {Giovanelli} R., {Haynes} M.~P., {Shankar} F., 2015, \aap,
  574, A113

\bibitem[{{Papastergis} {et~al}\mbox{.}(2011){Papastergis}, {Martin},
  {Giovanelli}, \& {Haynes}}]{papastergis11}
{Papastergis} E., {Martin} A.~M., {Giovanelli} R., {Haynes} M.~P., 2011, \apj,
  739, 38

\bibitem[{{Peeples} \& {Shankar}(2011)}]{peeples11}
{Peeples} M.~S., {Shankar} F., 2011, \mnras, 417, 2962

\bibitem[{{Pontzen} \& {Governato}(2012)}]{pontzen12}
{Pontzen} A., {Governato} F., 2012, \mnras, 421, 3464

\bibitem[{{Read} {et~al}\mbox{.}(2006){Read}, {Wilkinson}, {Evans}, {Gilmore},
  \& {Kleyna}}]{read06}
{Read} J.~I., {Wilkinson} M.~I., {Evans} N.~W., {Gilmore} G., {Kleyna} J.~T.,
  2006, \mnras, 367, 387

\bibitem[{{Schneider} {et~al}\mbox{.}(2014){Schneider}, {Anderhalden},
  {Macci{\`o}}, \& {Diemand}}]{schneider14}
{Schneider} A., {Anderhalden} D., {Macci{\`o}} A.~V., {Diemand} J., 2014,
  \mnras, 441, L6

\bibitem[{{Shankar} {et~al}\mbox{.}(2006){Shankar}, {Lapi}, {Salucci}, {De
  Zotti}, \& {Danese}}]{shankar06}
{Shankar} F., {Lapi} A., {Salucci} P., {De Zotti} G., {Danese} L., 2006, \apj,
  643, 14

\bibitem[{{Somerville}(2002)}]{somerville02}
{Somerville} R.~S., 2002, \apjl, 572, L23

\bibitem[{{Trachternach} {et~al}\mbox{.}(2009){Trachternach}, {de Blok},
  {McGaugh}, {van der Hulst}, \& {Dettmar}}]{trachternach09}
{Trachternach} C., {de Blok} W.~J.~G., {McGaugh} S.~S., {van der Hulst} J.~M.,
  {Dettmar} R.-J., 2009, \aap, 505, 577

\bibitem[{{Trujillo-Gomez} {et~al}\mbox{.}(2011){Trujillo-Gomez}, {Klypin},
  {Primack}, \& {Romanowsky}}]{tg11}
{Trujillo-Gomez} S., {Klypin} A., {Primack} J., {Romanowsky} A.~J., 2011, \apj,
  742, 16

\bibitem[{{Vogelsberger} {et~al}\mbox{.}(2012){Vogelsberger}, {Zavala}, \&
  {Loeb}}]{vogelsberger12}
{Vogelsberger} M., {Zavala} J., {Loeb} A., 2012, \mnras, 423, 3740

\bibitem[{{Zavala} {et~al}\mbox{.}(2009){Zavala}, {Jing}, {Faltenbacher},
  {Yepes}, {Hoffman}, {Gottl{\"o}ber}, \& {Catinella}}]{zavala09}
{Zavala} J., {Jing} Y.~P., {Faltenbacher} A., {Yepes} G., {Hoffman} Y.,
  {Gottl{\"o}ber} S., {Catinella} B., 2009, \apj, 700, 1779

\bibitem[{{Zwaan} {et~al}\mbox{.}(2010){Zwaan}, {Meyer}, \&
  {Staveley-Smith}}]{zwaan10}
{Zwaan} M.~A., {Meyer} M.~J., {Staveley-Smith} L., 2010, \mnras, 403, 1969

\end{thebibliography}
\appendix
\section{Different Data Sets}
We have emphasised in our paper the effect of the different measures of the BTFR on the velocity function. Our analysis adopts a set of empirical relations between \mhalo, \mstar,   M$_{\rm gas}$ and \vrot, using data from several different studies. 
Here, we repeat our analysis but starting with a different set of empirical relations between \mhalo, \mstar,  M$_{gas}$ and \vrot, using the results  of the recent study of \cite{bradford15}. who study HI line-widths of a sample of isolated dwarf galaxies selected from the Sloan Digital Sky Survey. We show that our main conclusions are not altered. 

\subsection{Different baryonic abundance matching}
\begin{figure}
\hspace{-.25cm}
\includegraphics[width=.43\textwidth]{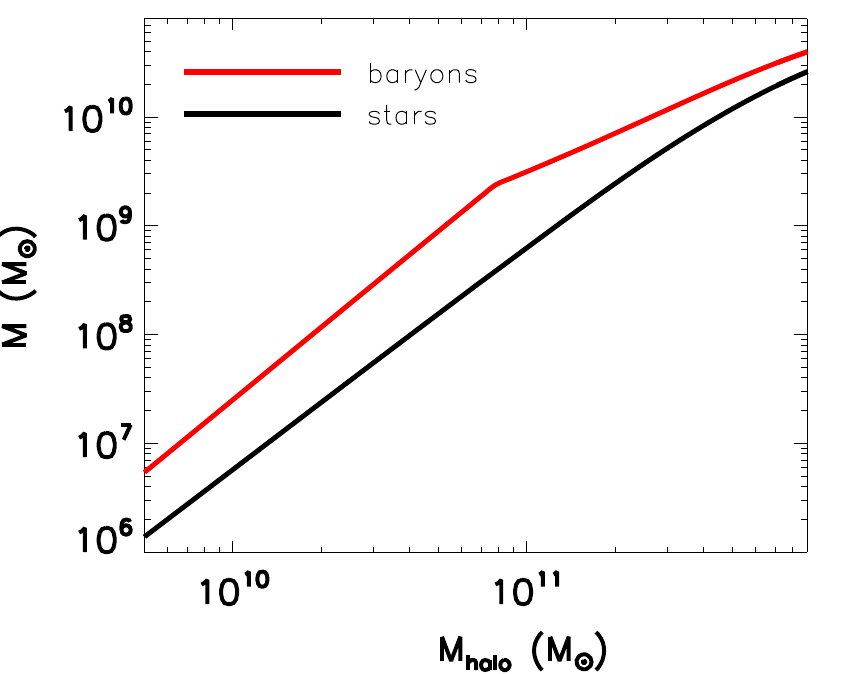}
\caption{The abundance matching relation adopted in the appendix, \mb-\mhalo\ (red line) and \mstar-\mhalo\ (black line). 
}
\label{A1}
\end{figure}

Our starting point is the empirical \mstar-\mhalo\ relation of \cite{moster10}, shown as a black line in figure~\ref{A1}, which is slightly flatter at the low mass end than the \cite{guo10} relation used in the paper.  We then use the double power law relation between \mstar\ and M$_{\rm gas}$ from \cite{bradford15}, which results in the   \mb-\mhalo\ relation shown as the red line in figure~\ref{A1}. 

\subsection{Reconciling BTFRs from a new data set}
\begin{figure}
\includegraphics[width=.43\textwidth]{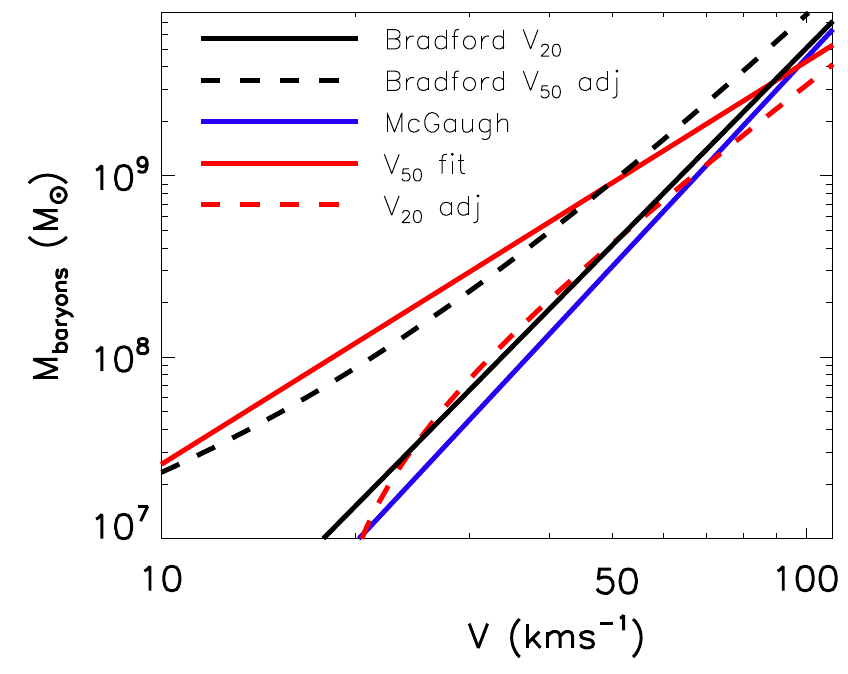}
\caption{The baryonic Tully-Fisher relation (BTFR) for the data set, with a fit to galaxies. Also shown are the BTFRs of McGaugh 2012, derived using \vflat.} 
\label{btfrA}
\end{figure}

\cite{bradford15} derive a BTFR using W20/2, 
\begin{equation}
 {\rm log}_{10}{\rm W}_{\rm 20}/2=-0.668+0.277{\rm log}_{10}{\rm M_b}
 \label{bradfordBTFR}
\end{equation}

\noindent where there is no correction made for dispersion. This BTFR is shown as a black line in Figure~\ref{btfrA}, and is is very similar to the BTFR from \cite{mcgaugh12}, shown as a blue line. We also show, as a red line, the \vfifty\ BTFR from this paper, i.e. equation~\ref{eq:w50btfr}. The dashed red line adjusts this \vfifty\ BTFR to a W20/2 BTFR using equation~\ref{eq:w50w20}, after also adjusting \vfifty\ to W50/2 by including the 8\,\kms velocity dispersion. The resultant relation is very close to the \vtwenty\ relation of \cite{bradford15}. 

\cite{bradford15} find a typical difference between W50 and W20 is $\sim$25\kms (see also \citealt{koribalski04}). We therefore use this value to also transform the \cite{bradford15} BTFR (black line) to a W50 BTFR, shown as the dashed black line. The adjusted BTFR is very similar to our adopted W50 BTFR, equation~\ref{eq:w50btfr}. This gives us significant confidence in our analysis, and in particular in equations equation~\ref{eq:w50btfr} \& \ref{eq:w50w20} from our paper, which are the key results which are driving the significant  differences  in the velocity function for different measures of \vrot.

\subsection{The effect of BTFRs on the Velocity Function}
\begin{figure}
\includegraphics[width=.43\textwidth]{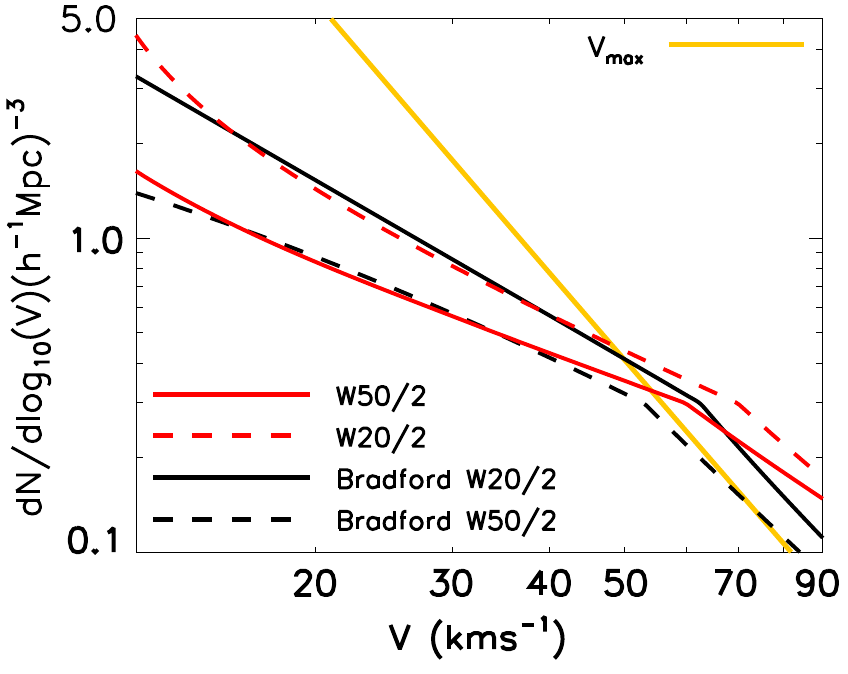}
\caption{The yellow line shows the V$_{\rm max}$ velocity function of DM only simulations. We then show the velocity function predicted by $\Lambda$CDM cosmology when making different assumptions about the BTFR, having matched observed baryon masses to DM halo masses. Using the W50/2 BTFR, equation~\ref{eq:w50btfr}, is shown as the red line. The W20/2 BTFR, derived using equation~\ref{eq:w50w20} is shown as a red dashed line. Results of assuming the BTFR of Bradford (2015), i.e. equation~\ref{bradfordBTFR}  are shown as the black solid line, with the adjustment of W50=W20-25\,\kms shown as the black dashed lines.}
\label{fig:VFnA}
\end{figure}

Starting this time from the abundance matching by mass shown in Figure~\ref{A1}, we  show in Figure ~\ref{fig:VFnA} the effect of the different BTFRs on the velocity function.
The yellow line shows the V$_{\rm max}$ velocity function of DM only simulations.  Using the W50/2 BTFR, equation~\ref{eq:w50btfr}, is shown as the red line. The W20/2 BTFR, derived using equation~\ref{eq:w50w20} is shown as a red dashed line. Results of assuming the BTFR of Bradford (2015), i.e. equation~\ref{bradfordBTFR}  are shown as the black solid line, with the adjustment of W50=W20-25\,\kms shown as the black dashed line.

%\bsp

\label{lastpage}

\end{document}